\documentstyle[12pt]{article}

\newcommand{\beg}{\begin{equation}}
\newcommand{\enq}{\end{equation}}
\newcommand{\sig}{\sigma}
\newcommand{\cst}{{\rm constant}}

\newcommand{\ap}{\alpha}
\newcommand{\ga}{\gamma}

\begin{document}

\title{Largest Eigenvalue Distribution in the Double Scaling Limit
of Matrix Models: A Coulomb Fluid Approach}
\author{Yang Chen$^{\dag}$, Kasper J.~Eriksen$^{\dag}$ and
Craig A.~Tracy$^{\dag\ddag}$\\
$^{\dag}$Department of Mathematics\\
Imperial College, London SW7 2BZ UK\\
$^{\ddag}$Department of Mathematics and Institute of Theoretical Dynamics\\
University of California, Davis, CA 95616,USA }
\date{\today}
\maketitle

\begin{abstract}
Using thermodynamic arguments we find that the probability  that
there are no eigenvalues in the interval $(-s,\infty)$
 in the double scaling limit of Hermitean matrix models is
${\rm O}\left(\exp(-s^{2\gamma+2})\right)$  as $s\rightarrow +\infty$.  Here
$\gamma=m-1/2$, $m=1,2,\cdots$ determine the $m^{{\rm th}}$
multi-critical point of the level density:
$\sig(x)\propto b\left[1-(x/b)^2\right]^{\gamma},\;x\in (-b,b)$, $b^2
\propto N$.
Furthermore, the size of the transition zone where the
eigenvalue density becomes vanishingly small at the tail of the spectrum
is $\approx N^{\frac{\gamma-1}{2(\gamma+1)}}$  in agreement
with earlier work based upon the string equation.
\end{abstract}
\par\noindent
\underline{To appear as a Letter in J. Phys. A: Math. Gen. 1995.}
\par\noindent
A basic quantity in Hermitean matrix models
is the probability,
$E_2(0;J)$, that a set $J$ contains no eigenvalues. For $N\times N$
Hermitean matrix models with unitary symmetry we have the well-known expression
\beg
E_2(0;J)=\frac{\int_{{\bar J}}{\rm e}^{-\sum_aV(x_a)}d\mu(x)}
{\int_{J\cup{\bar J}}{\rm e}^{-\sum_aV(x_a)}d\mu(x)}
=:\frac{Z[{\bar J}]}{Z[J\cup {\bar J}]}=:{\rm e}^{-\left(F[{\bar J}]-
F[J\cup{\bar J}]\right)},
\enq
with
$$d\mu(x)=\prod_{1\leq a<b\leq N}|x_a-x_b|^2\;\prod_{1\leq a\leq N}dx_a,$$
${\bar J}$ the complement of $J$ and
$V(x)$ is the ``confining'' potential\cite{Mehta}.
As indicated in (1), minus the logarithm of this probability
has the physical interpretation, in terms of Dyson's Coulomb fluid
\cite{Dy1,Dy2,Dy3,Chen}, as the change in free energy
\beg
\Delta F=F[{\bar J}]-F[J\cup{\bar J}];
\enq
that is, the free energy of the $N$ charges confined to region
${\bar J}$ minus the free energy of $N$ charges in the natural support
${J\cup{\bar J}}$ of $w(x):={\rm e}^{-V(x)}$.
\par
In this paper we shall mainly consider  the case $J=(-s,\infty)$, $s>0$, and
write $E_2(s)$ for $E_2(0;(-s,\infty))$.
Below we shall use
the {\it continuum approximation\/} of Dyson which treats the $N$ eigenvalues
in the large $N$ limit as a continuous fluid described by a continuous
charge density $\sig$ with the free energy expressed in terms of
$\sig $ \cite{Dy1}. This approximation has been previously applied to  the
Unitary Laguerre Ensemble (ULE) where
$w(x)=x^{\ap}{\rm e}^{-x}$,  $x\in(0,\infty)$ and $\ap>-1$
\cite{Chen}.\footnote{It is known from the theory of liquids
(by an application of
the Boltzmann principle) that the probability, $P_d(R)$, of finding
a bubble of radius $R$ in the bulk of a fluid (in $d$-dimensions) at
equilibrium with temperature $1/\beta$ is
$$P_d(R)\sim {\rm e}^{-\beta E_{V}R^{d}-\beta E_{\partial V}R^{d-1}},
\;\;R\gg {\rm coherence\;length},$$
where $E_V$ is the energy/volume for creating a bubble and $E_{\partial V}$
is the surface energy. If we specialise this formula to $d=1$ then
$$P_1(R)\sim {\rm e}^{-{\rm constant}R},$$ in contradiction
with the known result\cite{Dy1}. This is due to the fact the Coulomb fluid
has long ranged interactions.}
Here we examine
matrix models with\footnote{The reason for our
choice of notation for the coefficients
of $V$ will become clear below.}
\beg
V(x)=\sum_{k=0}^{p}\frac{g_{2k+2}}{(k+1)\;b^{2k}}\;x^{2k+2},
\label{V}\enq
with $g_2=1$.
In principle we should not have to make the continuum approximation since it is
known that $E_2(0;J)$ is
expressible in terms of solutions to a completely integrable system
of partial differential equations \cite{Tracy3}.
However, the {\it analysis\/}
of these equations for $V$ of the above form is quite difficult.  (Of
course, the Gaussian case is not included in this remark.)\ \ It
is hoped that the approximate expressions derived here, which we believe
are asymptotic as $s\rightarrow \infty$,  will aid in the analyses of these
equations.
\par
To begin,   consider
the Gaussian Unitary Ensemble (GUE) with $g_{2k+2}=0$ for $k\geq 1$.
For the scaled GUE with $J=(-t,t)$
it is known that
$E_2(0;(-t,t))$ is a  $\tau$-function of a
particular fifth Painlev\'e transcendent
\cite{Jimbo}.   Starting with this representation an
asymptotic expansion for $E_2(0;(-t,t))$ as $t\rightarrow\infty$
can  be derived, though the first such asymptotic expansion
was achieved by Dyson using methods of inverse
scattering \cite{Dy4}.  (Actually, there is still an undetermined constant
from either the inverse scattering analysis or the
 Painlev\'e analysis, see, e.g.~\cite{basor}.)
 The leading term, $-\ln E_2(0;(-t,t))\sim \frac{1}{2}\pi^2t^2$,
was first obtained from the fluid approximation\cite{Dy1}.
Indeed, the $t^2$ term of the asymptotic expansion can be
given a simple physical interpretation: it is proportional to the
square of the number of eigenvalues excluded in
the (scaled) interval $(-t,t)$ since in the bulk scaling limit of the GUE
the eigenvalue density is a constant $\sim\frac{\sqrt {2N}}{\pi}$.
This suggests that a natural variable  is one which
gives uniform density in the excluded interval.
 We can  always achieve this by a simple  change of variables
 since the problem is  one-dimensional.
By introducing a new variable $\xi$
and a corresponding $\varrho(\xi)$ via the relation
\beg
\varrho(\xi)d\xi:=1\cdot d\xi=\sig(x) dx,
\enq
the density in the $\xi$ ``scale'' is made unity. Therefore,
$-\ln E_2(0;J)$ is asymptotic to
$$\left[\int_{\xi_1}^{\xi_2}d\xi\right]^2=
\left[\int_{x_1}^{x_2}\sig(x)\;dx\right]^2,\;\;J=(x_1,x_2).
$$
We conclude from the above arguments that for a large interval,
\beg
-\ln E_2(0;J)\sim N^2(l),
\label{result1}\enq
where
\beg
N(l)
={\rm number\;of\;eigenvalues\;excluded\;in\;an\;interval\;of\;length}
\;l.
\label{result2}\enq
 We mention that a
  screening theory of the continuum  Coulomb fluid gives a physical
justification of these
arguments \cite{Chen1} though we know of no proof of the general validity
of this relationship.
\par
To further test the validity of (\ref{result1}) and
(\ref{result2} ) we consider the edge
scaling limit of the GUE where exact results are known \cite{Tracy1}.
Accordingly, we simply compute the number of eigenvalues excluded
from an interval of length $l\;(=b-a)$ from the soft edge,
$b={\sqrt {2N}}$,
$$
N(l)=\int_{a}^{b}dx\frac{1}{\pi}{\sqrt {b^2-x^2}}
\propto {\sqrt {2b}}\int_{a}^{b}dx{\sqrt {b-x}}\propto
\left[2^{1/2}N^{1/6}l\right]^{3/2}=:s^{3/2}.
$$
Observe that $N^2(l)\sim s^{3}$ not only supplies the correct exponent
in the scaled variable $s\; (=2^{1/2}N^{1/6}l)$  in $-\ln E_2(s)$,  we
also have the correct density at the soft edge:
$\sig_N({\sqrt {2N}})=2^{1/2}N^{1/6},$
which agrees with known exact results \cite{Mehta, Tracy1}.
This result predicts the shrinking of the
size of the transition zone ($\sim N^{-1/6})$ as
$N\rightarrow \infty$---a reasonable behaviour from
the Coulomb fluid point of view since the
GUE potential $x^2$ is strongly confining. The same approximation
has been applied to the origin scaling limit
of the ULE \cite{Chen}
and the result  agrees with the first
term of the exact asymptotic expansion \cite{Tracy2}.
\par
These two confirmations of the validity of (\ref{result1}) and (\ref{result2})
 give us confidence to apply the method to the
 matrix models  with $V$ given by (\ref{V}). (These are
 the cases of interest in the matrix models of 2D quantum
 gravity \cite{Bre1, Gross}.)\ \
 The charge density
$\sig$ satisfies an integral equation\cite{Dy1,Dy2} derived from the
following minimum principle:
$${\rm min}_{\sig}F[\sig],$$
\beg
F[\sig]=\int_{J}dxV(x)\sig(x)-\int_{J}dx\int_{J}dy\sig(x)\ln|x-y|\sig(y)
\enq
subject to the constraint $\int_{J}dx\sig(x)=N,$ which is
\beg
V(x)-2\int_{-b}^{b}dy\ln|x-y|\sig(y)=\cst={\rm chemical\;potential},
\;\;x\in(-b,b).
\label{int1}\enq
Since $V$ is even  so is $\sig$, and making use of this symmetry
(\ref{int1}) becomes
\beg
V(x)-2\int_{0}^{b}dy\ln|x^2-y^2|\sig(y)=\cst.
\label{int2}\enq
With the change of variables $x^2=u$,  $y^2=v$ and
$r(u)=\sig({\sqrt u})/(2\sqrt{ u})$,   (\ref{int2}) becomes
\beg
V({\sqrt u})-2\int_{0}^{b^2}dv\;r(v)\ln|u-v|=\cst,\;\;u\in(0,b^2).
\enq
This is converted into a singular integral equation by
differentiating with respect to $u$:
\beg
\frac{dV({\sqrt u})}{du}-
2\;{\rm P}\int_{0}^{b^2}dv\frac{r(v)}{u-v}=0,\;\;u\in (0,b^2).
\label{int3}\enq
 Here $b$, which determines the upper and
lower band edges, is fixed by the normalization condition
$\int_{-b}^{b}\sig(x)dx=N$.
\par
Following \cite{Akhiezer}
the solution is\footnote{$\frac{\cst}{{\sqrt {u(b^2-u)}}}$ solves
the homogeneous part of (\ref{int3}). However, based on the variational
principle, including this
solution would increase the free energy. $\;_2F_1(-k,1,3/2,z)=
\sum_{n=0}^{k}\frac{(-k)_n}{(3/2)_n}
\;z^n$ is a polynomial of degree $k$ in $z$.}
$$
r(u)=\frac{1}{2\pi^2}{\sqrt {\frac{b^2-u}{u}}}\;{\rm P}\int_{0}^{b^2}
\frac{dv}{v-u}{\sqrt {\frac{v}{b^2-v}}}\frac{dV({\sqrt u})}{du},
\;\;u\in (0,b^2)
$$
\beg
={\sqrt {\frac{b^2-u}{u}}}
\sum_{k=0}^{p}
t_k\;_2F_1\left(-k,1,\frac{3}{2},1-\frac{u}{b^2}\right),
\enq
where the integral can be found in \cite{Gr}
and
$$
t_k:=-\frac{1}{2\pi^2}B\left(-\frac{1}{2},k+\frac{3}{2}\right)g_{2k+2}.
$$
Returning  to $\sig$, it can be shown that
\beg
\sig(x)=b{\sqrt {1-\left(\frac{x}{b}\right)^2}}\;\;
\Pi_{p}\left[\left(\frac{x}{b}\right)^2\right],
\enq
where
$\Pi_p(z)$ is a polynomial of degree $p$ in $z$
with coefficients depending on the
linear combinations of the coupling constants $g_k$. The edge parameter
$b$
is determined from the normalization condition and reads $b^2=CN$ where
$$C=\frac{1}{\int_{-1}^{+1}dt{\sqrt {1-t^2}}\;\Pi_p(t^2)},$$
is independent of $N$.
\par
Taking the special case $p=1$ (now $g_4=g$),
we have
$$\sig(x)=\frac{b}{\pi}{\sqrt {1-\left(\frac{x}{b}\right)^2}}
\left[1+\frac{g}{2}+g\left(\frac{x}{b}\right)^2\right].
$$
By tuning $g$ to $g_c$, such that $-g_c=1+g_c/2$,  we have
$$\sig(x)=\cst\;b\left[1-\left(\frac{x}{b}\right)^2\right]^{3/2},
$$
producing a qualitative deviation in the density at the edges
$(\pm b)$ of the spectrum from the Wigner's
semi-circle distribution \cite{Bre3, Gross}.
A  calculation now gives
\beg
N(l)\propto \int_{a}^{b}dx\;b(1-x/b)^{3/2}(1+x/b)^{3/2}\approx
\frac{b}{b^{3/2}}\int_{a}^{b}(b-x)^{3/2}\sim
\left(\frac{l}{N^{1/10}}\right)^{5/2}=:s^{5/2},
\enq
and thus
$-\ln E_2(l)\sim s^{5}$.  Observe that due to this tuning  the length of the
transition zone ($\sim N^{1/10}$) is now an increasing function of $N$.
It is clear that the tuning procedure can be generalized to $p>1$\cite{Bre1}.
By simultaneously adjusting the coupling constants $g_4$, $g_6$ etc.,
to their respective critical values we can have
\beg
\sig(x)=\cst\;b\left[1-\left(\frac{x}{b}\right)^2\right]^{\ga},
\enq
where
$\ga=p+\frac{1}{2}$.\footnote{In the quantum gravity literature
$\ga=m-1/2,\;m=1,2\cdots$.} Computing $N(l)$ we find,
\beg
N(l)\propto \int_{a}^{b}dx\;b\left(1-\frac{x}{b}\right)^{\ga}
\;\left(1+\frac{x}{b}\right)^{\ga}\propto
\left(\frac{l}{N^{\frac{\ga-1}{2(\ga+1)}}}\right)^{\ga+1}
=:s^{\ga+1}.
\enq
Therefore
\beg
\log E_2(s) \approx -s^{2\ga+2},\;\;(s\rightarrow \infty).
\enq
The non-perturbative soft edge density is determined as
\beg
\sig_N({\sqrt N})\approx N^{\frac{1-\ga}{2(1+\ga)}},\;\;
N\rightarrow \infty.
\label{sigmaResult}\enq
The corresponding size of the transition zone is
$\approx N^{\mu}$,  where
$$\mu=\frac{\ga-1}{2(\ga+1)}$$
a result  previously obtained from the string equation \cite{Bre3,Fran}.
Note that our $x$ variable
is related to Bowick and Br\'ezin's \cite{Bre3} $\lambda$ as $x={\sqrt
N}\lambda$.
Supplying the appropriate ${\sqrt N}$ factor we obtain from (\ref{sigmaResult})
Bowick and Br\'ezin's result $N^{-\frac{2}{2m+1}}$.
\vskip 1cm
\noindent
{\bf Acknowledgements}\\
\noindent
The authors wish to
 thank Dr.~Di Francesco for providing us his  unpublished notes \cite{Fran}.
 In addition one of us (Y.~C.)  thanks Professor C.~Itzykson
for discussing the problem of level-spacing distribution in this model.
Finally third author gratefully acknowledges the Science and Engineering
Research
Council, UK for the award of a Visiting Fellowship,
the National Science Foundation through grant DMS--9303413, and the Mathematics
Department of Imperial College for its kind hospitality.
\vfill\eject


\begin{thebibliography}{99}
\bibitem{Akhiezer} N.~I.~Akhiezer and I.~M.~Glazman, {\em
Theory of Linear Operators In Hilbert Space} Trans.\ Merlynd Nestell
vol.~1 (New York: Ungar) 1961.

\bibitem{basor} E.~L.~Basor, C.~A.~Tracy, and H.~Widom, Phys.\ Rev.\ Letts.\
{\bf 69} (1992) 5; H.~Widom,   J.\ Approx.\ Th.\ {\bf 76}
(1994) 51.


\bibitem{Bre1}E.~Br\'ezin in
{\em Two Dimensional Quantum Gravity and Random Surfaces}
pg.1, Eds., D.~J.~Gross, T.~Piran and S.~Weinberg, World Scientific Publ. 1992.

\bibitem{Bre2}E.~Br\'ezin, C.~Itzykson, G.~Parisi and J.~B.~Zuber,
Commun.\  Math.\  Phys.\ {\bf 59} (1978) 35.

\bibitem{Bre3} M.~J.~Bowick and E.~Br\'ezin, Phys.\ Lett.\  {\bf B268}
(1991) 21.

\bibitem{Chen}Y.~Chen and S.~M.~Manning, J.\  Phys.\  A {\bf 27} (1994) 3615.

\bibitem{Chen1}Y.~Chen and K.~J.~Eriksen, {\em Level Spacing Distribution of
The $\ap-$  Ensemble}, (1994) preprint.

\bibitem{Fran}P.~Di Francesco, unpublished notes, 1994.

\bibitem{Dy1}F.~J.~Dyson, J.\ Math.\  Phys.\  {\bf 3} (1962) 157.
\bibitem{Dy2}F.~J.~Dyson, J.\  Math.\  Phys.\  {\bf 13} (1972) 90.
\bibitem{Dy4}F.~J.~Dyson, Commun.\ Math.\ Phys.\ {\bf 47} (1976) 171.
\bibitem{Dy3}F.~J.~Dyson, {\em The Coulomb Fluid And The fifth
Painlev\'e Transcendent} IASSNAA-HEP-p2/43 1992 preprint to appear
in Proc. Conf. in honour of C.~N.~Yang.

\bibitem{Gr}I. S.~Gradshteyn and I.~M.~Ryzhik,
{\em Table of Integrals, Series and Products}, Academic, London 1980,
formula 3.2283.

\bibitem{Gross}D.~J.~Gross and A.~A.~Migdal, Phys.\ Rev.\  Letts.\
{\bf 64} (1990) 127;  M.~R.~Douglas and S.~H.~Shenker, Nucl.\  Phys.\
{\bf B335} (1990) 635;  E.~Br\'ezin and V.~A.~Kazakov, Phys.\  Letts.\
{\bf B236} (1990) 144.

\bibitem{Jimbo}M.~Jimbo, T.~Miwa, Y.~M\^ori and M.~Sato, Physica
{\bf 1D} (1981) 407. See also, C.~A.~Tracy and H.~Widom,
{\em Introduction to Random Matrices,} in {\em Geometric and Quantum Aspects
of Integrable Systems,} G.~F.~Helminck, ed., Lecture Notes in Physics,
Vol. 424, Springer-Verlag (Berlin), 1993, pgs.\ 103--130.

\bibitem{Mehta}M.~L.~Mehta, {\em Random Matrices},
2$^{{\rm nd}}$ Ed., Academic Press (New York) 1991.

\bibitem{Tracy1}C.~A.~Tracy and H.~Widom, Commun.\  Math.\ Phys.\
{\bf 159} (1994) 151.
\bibitem{Tracy2}C.~A.~Tracy and H.~Widom, Commun.\  Math.\  Phys.\
{\bf 161} (1994) 289.
\bibitem{Tracy3}C.~A.~Tracy and H.~Widom, Commun.\  Math.\  Phys.\
{\bf 163} (1994) 33.
\end{thebibliography}
\end{document}